\documentclass[journal=jpclcd,manuscript=letter]{achemso}

\usepackage{chemformula} 
\usepackage[T1]{fontenc} 
\usepackage{float}
\usepackage{subfig}

\author{David Ribar}
\affiliation{Computational Chemistry, Lund University, P.O.Box 124, S-221 00 Lund, Sweden}
\author{Clifford E. Woodward}
\affiliation{School of Physical, Environmental and Mathematical Sciences University College, University of New South Wales, ADFA Canberra ACT 2600, Australia}
\author{Sture Nordholm}
\affiliation{Department of Chemistry and Molecular Biology, The University of Gothenburg, 412 96 Gothenburg, Sweden}
\author{Jan Forsman}
\email{jan.forsman@compchem.lu.se}
\affiliation{Computational Chemistry, Lund University, P.O.Box 124, S-221 00 Lund, Sweden}

\title{Cluster Formation induced by local dielectric saturation in Restricted Primitive Model Electrolytes}


\begin{document}

\begin{tocentry}
\includegraphics[scale=1]{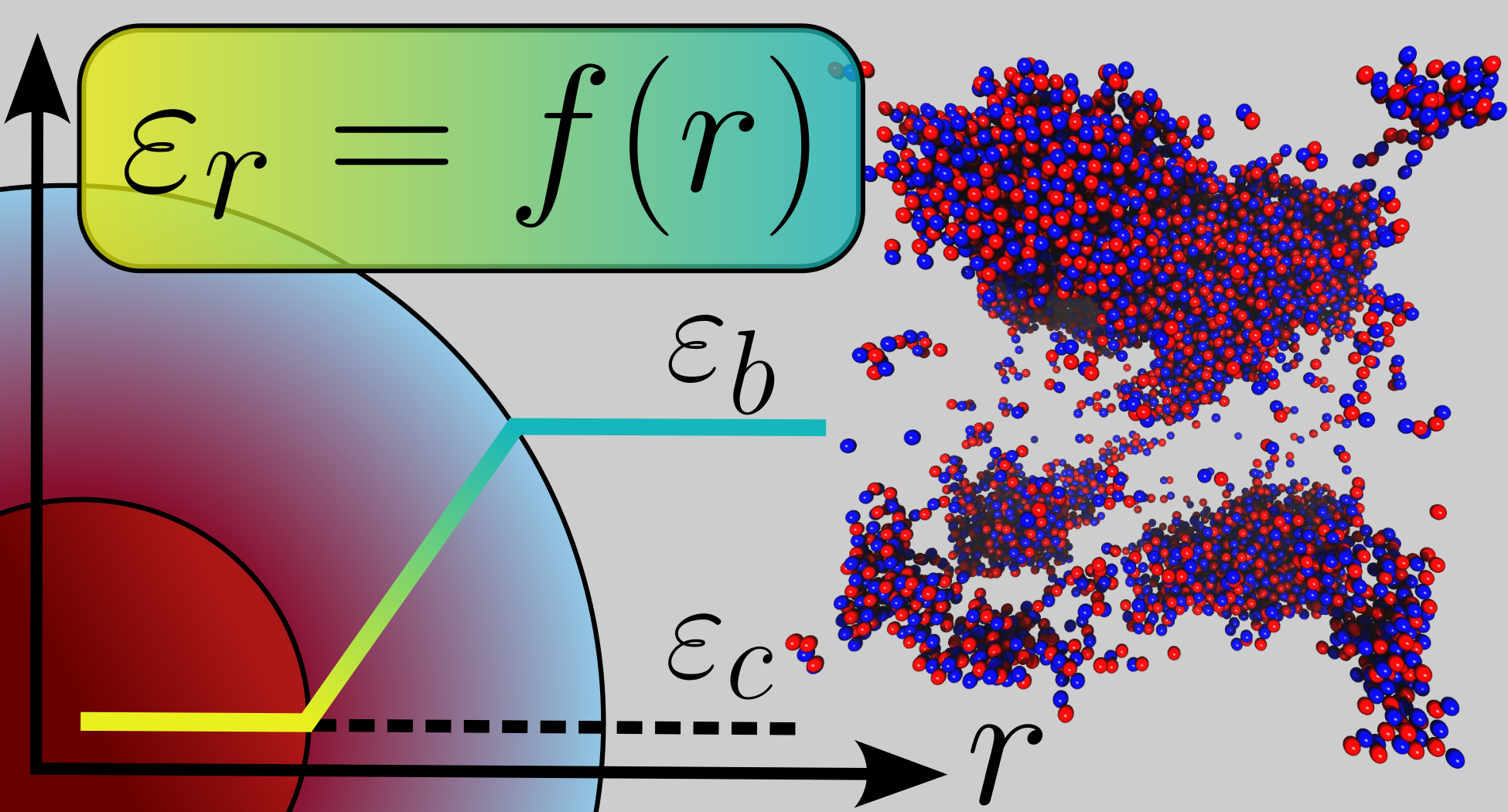}  
\end{tocentry}

\begin{abstract}
  Experiments using the Surface Force Apparatus (SFA) have found anomalously long ranged charge-charge underscreening in concentrated salt
  solutions. Meanwhile, theory and  simulations have suggested ion clustering to be the possible origin of this behaviour. The popular
  Restricted Primitive Model of electrolyte solutions, in which the solvent is represented by a
  uniform relative dielectric constant, $\varepsilon_r$, is unable to resolve
 the anomalous underscreening seen in experiments. In this work, we modify the Restricted Primitive Model to account
  for local dielectric saturation within the ion hydration shell.
  The dielectric constant in our model locally decreases from the bulk value to a lower saturated value at the ionic surface.
  The parameters for the model are deduced so that typical salt solubilities are obtained. 
  Our simulations for both bulk and slit geometries show that our model displays
  strong cluster formation and these give rise to long-ranged interactions between charged surfaces
  at distances similar to what has been observed in SFA measurements. An electrolyte
  model wherein the dielectric constant remains uniform does not display similar clusters, even with $\varepsilon_r$ 
  equal to the saturated value at ion contact.  Hence, the observed behaviours are not simply due to an enhanced Coulomb interaction. 
\end{abstract}

\section{}

Electrolytes play an important role in a plethora of both scientific and industrial applications \cite{Israelachvili:91,Evans:94,Holm:01}.
Simple theoretical descriptions at the mean-field level \cite{Derjaguin:41,Verwey:48,Israelachvili:91} have proven to be reasonably
accurate for aqueous electrolytes at low coupling strength, e.g., monovalent ions (1:1 salts) in aqueous solvents at low and intermediate concentrations.
Perhaps the most fundamental prediction by these theories is so-called {\em ionic screening}, often described in terms of 
the so-called Debye screening length, $\lambda_D$.  The Debye length is predicted to be inversely proportional
to the square root of the electrolyte concentration, $\lambda_D \sim 1/c^{1/2}$.  Salts composed of ions where at least one
component is multivalent will generally
require a higher level of theory, accounting for ion-ion correlations \cite{Nordholm:84a,Guldbrand:84,Kjellander:86,Torrie:91}. 
This notwithstanding, it was believed that mean-field theories could capture the qualitative behaviours
of even concentrated aqueous solutions of 1:1 salts, with some predictable corrections due to correlations. 
This has been called into question by recent experimental investigations of 1:1 electrolyte solutions, 
above a threshold concentration of about 1 M.  These have revealed a
peculiar anomalous underscreening phenomenon \cite{Gebbie:2013, Gebbie:2015, Smith:2016, D3FD00042G, C6FD00250A,Yuan:22}
whereby the interaction between charged surfaces exhibit an exponential decay with an extraordinarily long range, $\lambda$,
much larger than the Debye length predicted by mean-field theory.   
Moreover, $\lambda$  appears to {\em increase} with the salt concentration, in qualitative disagreement with
more sophisticated theories that attempt to correct for correlations \cite{Attard_1993, coupette_screening_2018,Hartel:23}.
It should be noted that anomalous underscreening has been experimentally challenged \cite{Kumar:22}, and despite considerable
theoretical efforts \cite{coupette_screening_2018,Rotenberg:18,Kjellander:20,Coles:20,Zeman:20,Cats:21,Hartel:23} there is at present no consensus
as to its physical origin.

One often proposed explanation involves the role of clusters and their 
effect on the Debye screening length. For example, in a concentrated  1:1 electrolyte, 
the formation of neutral and/or weakly charged clusters in concentrated ionic solutions would act to reduce the number
of independent charged species. If, $c*$, denotes the concentration of the {\em effective} screening charge (and we assume  
most clusters are either neutral or univalent), then the modified Debye length would be given by,  $\lambda \sim 1/c*^{1/2}$.
Screening length measurements in the anomalous region have been fit to the relation,
$\lambda \sim c l_B d^3$, where $l_B = \beta e_0^2 /(4\pi\varepsilon_0\varepsilon_r)$  is the Bjerrum length and $d$ is
an average ionic diameter.  Here $\beta = (k_BT)^{-1}$ is the inverse thermal temperature,  $e_0$ denotes the elementary
charge, and $\varepsilon_0$ is the the permittivity of vacuum.
Attributing underscreening to clustering and a modified Debye length, would require 
$c^* \sim 1/c^2$ in the anomalous region.

It is worthwhile to reflect upon the implications of such an explanation in a little more detail.  
Suppose, we consider an electrolyte solution with a concentration well below the threshold value 
($\sim$ 1 M) above which underscreening occurs.  Presumably, there will be some 
incipient clustering of ions at this concentration, but insufficient to affect the screening 
length significantly.  Clustering at low concentration is driven by the electrostatic attractions between ions,
but clusters remain finite due to the low chemical potential of the free ions (with which the aggregated ions
remain in equilibrium).  As the concentration increases, clustering is additionally aided by the
lowering of the overall excluded volume due to ion aggregation.  At concentrations
lower than the threshold value for underscreening, addition of more ions 
will tend to lower the screening length, as the number of free ions
at equilibrium will generally increase with the overall concentration.  
Once the concentration reaches the underscreening threshold value, however, 
addition of more ions will cause the screening length to increase.  Since we are now in the region
of anomalous underscreening, effectively {\em all} the added ions will be aggregated to create {\em neutral} clusters. 
In addition, a portion of the original population of charged species (charged clusters and free ions) must also  
reorganise into neutral structures. If correct, this has the hallmarks of an apparent instability in the free energy of cluster formation.  
It suggests that, at a threshold concentration of free ions, clusters become unstable and experience accelerated growth, 
presumably because they have surpassed a critical size. This growth is arrested by a concurrent decrease in the 
free ion concentration.  Experiments indicate that the concentration of effective screening charges would 
need to be some $10^4$ times lower than $c$ in order to explain the upper levels of
underscreening \cite{Gebbie:2013, Ma:2015, Gebbie:2015, Smith:2016}.

In electrolyte theory the  Restricted Primitive Model (RPM) has been something of a "workhorse".  
Here ions are treated as charged hard spheres with diameter, $d$, and the solvent is simply modelled using a 
uniform dielectric constant, $\varepsilon_r$, which is generally greater than 1 due to the polarisability 
of solvent molecules.  Thus, ion-ion interactions are described by,  $\phi_{ij}(r)$,  where: 
\begin{equation}
		\beta\phi_{ij}(r) = \left\{
	\begin{array}{ll}
		\infty ; & r \leq d \\
		l_B \frac{z_i z_j}{r}; & r > d  \\
	\end{array}
	\right.
	\label{eq:coulomb}
\end{equation}
Here, $z_i$ and $z_j$ denote the valencies of interacting ions $i$ and $j$.
Interestingly, 
hard sphere interactions tend to shorten the electrostatic screening length compared with $\lambda_D$,  due to interplay 
between hard sphere and electrostatic correlations. \cite{Attard_1993}.  

A clustering mechanism was recently explored using simulations of the 
RPM\cite{Hartel:23}. The simulations showed that the propensity of clusters to form increases 
when the coupling strength and/or the electrolyte concentration increases and that this does lead to 
underscreening.    Furthermore, a clustering analysis was used to confirm that the measured screening length 
was consistent with using the apparent concentration of independent charge carriers, $c*$
(charged clusters and free ions), i.e., $\lambda \sim 1/c*^{1/2}$.  However, the underscreening
observed was insufficient to explain experimental results. In particular, the
simulations did not predict an increase in the observed screening length as the 
overall electrolyte concentration increases, i.e., the simulated $c*$ {\em always increased}
with $c$.   It should be noted that these simulations also accounted for a reduction in $\varepsilon_r$
with increasing $c$.   Thus the RPM is not able to reproduce the trends
seen in experiments, which is perhaps not unanticipated given the discussion above.
That is, the RPM appears to have no inherent features that would suggest the possibility of 
anomalous clustering of the type discussed above.  

Given this, it is interesting to consider other explanations (apart from clustering) 
to explain anomalous underscreening.  In particular,
we consider here the possibility that the decay length in experiments may be a reflection of the
{\em size} of clusters, rather than their impact on the concentration of free charges.
In the case of NaCl, the largest decay lengths measured in the anomalous
regime are of the order of $30$ \AA. \cite{Smith:2016}  A pair of $Na^+$ $Cl^-$ ions in contact has a size of $\sim 6$ \AA, 
while water molecules have a diameter $\sim 3$ \AA.  Thus, one could envisage a small cluster
of a few pairs of ions and hydrating water molecules to have a size of at least  $\sim 30$ \AA. 
While such a mechanism has not been uncovered for the RPM, that may be a consequence of short-comings
in that interaction model, which does not allow the formation of large enough clusters.

Most simple theories and simulation models of electrolytes do not explicitly include the solvent
but allows it to be represented in a ``primitive'' fashion, via a relative dielectric constant, $\varepsilon_r$, which is
greater than unity due to the
interaction of the ion charge with electrons and nuclear charges of the surrounding solvent molecules. Water is a polar
solvent and can reorient and rearrange in response to the electric field of an ion. Experimentally, it is known that as
the concentration of an aqueous electrolyte solution increases the 
overall dielectric constant
decreases \cite{hasted_dielectric_2004, deSouza2022, Conway83, Bonthuis2012, DanielewiczFerchmin2013, adar_dielectric_2018}. 
This can be rationalised if we consider that hydrating water molecules are rotationally constrained
by electric fields from the ions.  As the salt concentration increases to molar levels,
these constrained water molecules constitute an increasingly significant percentage of the solvent, which 
leads to a decreased overall dielectric response.  
This type of {\em dielectric saturation} has been the focus of some previous investigations at
electrode surfaces, or in confined geometries\cite{Bonthuis2012, deSouza2022, Underwood2022}.
Theoretical investigations have explored this reduction in
dielectric permittivity by correcting the mean field approach in a consistent manner, \cite{Kjellander:20}
or else model it in terms of ion-specific effects \cite{ben-yaakov_dielectric_2011, hubbard_molecular_1979}.
In protein small angle x-ray scattering model fitting procedures, the formation of water hydration shells around charged
(and uncharged) proteins, is a well-established phenomenon. A special treatment of hydration shells is
needed, in order to produce a model fit that correctly describes the measured spectra \cite{Knight2015, Hansen2021, Hansen2022}. 

We propose to modify the RPM, based on physically plausible arguments.  
To maintain simplicity we will explore a 2-body interaction
model, but ultimately the effects we consider are best manifested using many-body forces.  
This notwithstanding, we consider it useful to understand the effect on screening lengths in the presence of 
much larger clusters than can be generated by the simple RPM.  We use this model to roughly mimic 
NaCl and show that it saturates at a concentration close to its observed 
value, if not somewhat lower.  While we by no means present this 
new model as an accurate representation of aqueous NaCl solutions, it does allow
us to investigate screening behaviour in electrolytes
in the presence of much larger clusters than is generated by the RPM.  This allows us
to gain some further insight as to potential mechanisms that a cluster model can provide 
to explain anomalous underscreening apart from the usual assertion that it reduces 
the concentration of free charges.     

A physically reasonable modification of the RPM, to account for dielectric saturation, would be to make  
$\varepsilon_r$ in Eq.(\ref{eq:coulomb}) a function of the average 
electrolyte concentration, as has been employed in previous RPM studies. \cite{Hartel:23}. This will increase
the coupling strength between ions at higher average electrolyte concentration.    
However, even in dilute solutions, we expect that dielectric saturation will occur in regions 
where fluctuations cause a locally high concentration of ions.  In such a region, 
solvent molecules will be displaced, and those that remain are subject to
large electric fields \cite{Bonthuis2012}.  
This suggests a spatial inhomogeneity in $\varepsilon_r$ should occur as a 
result of changes in the ionic configurations.  In principle, dielectric inhomogeneity 
should be modelled as a many-body phenomenon that qualitatively results in  
local changes to the electrostatic interactions.  A collection of ions will produce large fields
reducing the local dielectric constant and promoting clustering. Countering this will be an 
energetically unfavourable contribution  due to the overlap of solvation shells of those ions.  
In this work, we consider a phenomenological manifestation of these mechanisms at the 
pair-wise interaction level, in order to maintain computational simplicity.

In the model used here, $\varepsilon_r$, is assumed to vary with the ionic separation, $r$.  
Specifically, it will have a reduced value when ions approach each other but has its {\em bulk} value
when the ions are sufficiently separated.  We will model  $\varepsilon_r(r)$ as a linear ramp function of the following form:
\begin{equation}
	\varepsilon_r(r) = \left\{
	\begin{array}{ll}
		\varepsilon_c ; & r \leq d \\
		\varepsilon_c + (\varepsilon_b - \varepsilon_c) \frac{r - d}{\Delta}; & d < r \leq d + \Delta  \\
		\varepsilon_b; & r > d + \Delta  \\
	\end{array}
	\right.
	\label{eq:vareps_def}
\end{equation}
\noindent with $\varepsilon_c, \varepsilon_b$ denoting the \textit{contact} and \textit{bulk solvent} 
dielectric constant values, respectively.
In general we have  $\varepsilon_c$ <  $\varepsilon_b$
The slope of the linear ramp is defined by the parameter $\Delta$, 
which is taken as the diameter of a solvent molecule, $d$.  We chose a value of $d=$ 3 Å  representing
a single hydration layer of water molecules.  Consistent with this, we set $\varepsilon_b = 78.3$.
The thus {\em modified} potential used here, consists essentially of a short-ranged interaction 
(attractive between unlike ions, repulsive between like ions) in 
addition to a typical RPM interaction, wherein the latter  
assumes the uniform dielectric constant of the bulk solvent, $\varepsilon_b$.
While other implementations of the RPM have assumed a uniform  dielectric constant, $\varepsilon_r$,
that decreases with electrolyte concentration to account dielectric saturation,     
we shall see below that a locally dependent $\varepsilon_r$ of the type proposed here
promotes more clustering, even compared to an RPM that uses $\varepsilon_r = \varepsilon_c$ everywhere.
We will use our potential model to explore both bulk simulations of the electrolyte as well as 
the solution in contact with charged  surfaces.      

We will only give a brief account of the simulation methods here, and refer to
the Supplementary Information (SI) for details.
Canonical ensemble Metropolis Monte-Carlo simulations were performed
at 298 K for two geometries: a cubic simulation geometry, henceforth referred to as the {\em bulk}, 
and a parallelepiped geometry between two charged hard walls, referred
to as the {\em slit}. Periodic boundary conditions were applied along $(x,y,z)$ for the bulk 
simulations (with side-length of the cubic simulation box denoted as $L$) and
along $(x,y)$ for the slit simulations, with impenetrable charged hard walls situated at $z=\pm H/2$. 
For slit systems, $L$ denotes the length of the simulation
parallelepiped along the $(x,y)$-axis, and $H$ represents the slit width along the $z$-axis. 
The values of $L$ and $H$ are adjusted to accommodate the
different electrolyte concentrations under investigation. For the bulk simulations, 
we employed Minimum Image (MI) truncation of the Coulomb interactions. 
The accuracy of this choice has been evaluated thoroughly in a recent work \cite{Forsman:24a}, 
by direct comparisons with more elaborate (and computationally expensive) Ewald simulations. 
These comparisons verified that MI truncation leads to structurally accurate results for bulk systems.
We have included yet another comparison, with the same conclusion, in the SI.
For the slit systems, we employed the standard ``charged sheet'' method \cite{Torrie:80} to
account for long-ranged interactions. Notably, cluster moves were implemented for 
both systems, leading to crucial improvements of the statistical performance.

\begin{figure}[h!]
  \centering
  \subfloat[]{
    	\includegraphics[scale=0.36]{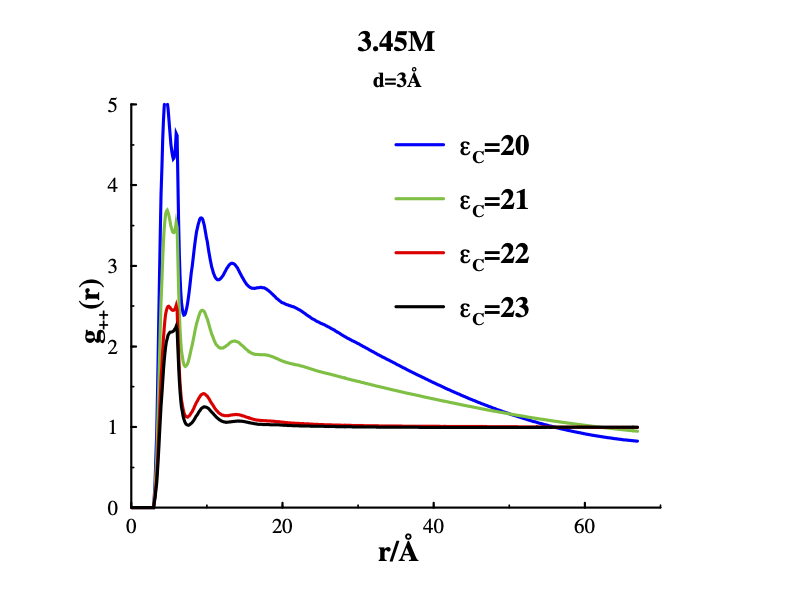}
    }
    \hfill
    \subfloat[]{
	\includegraphics[scale=0.36]{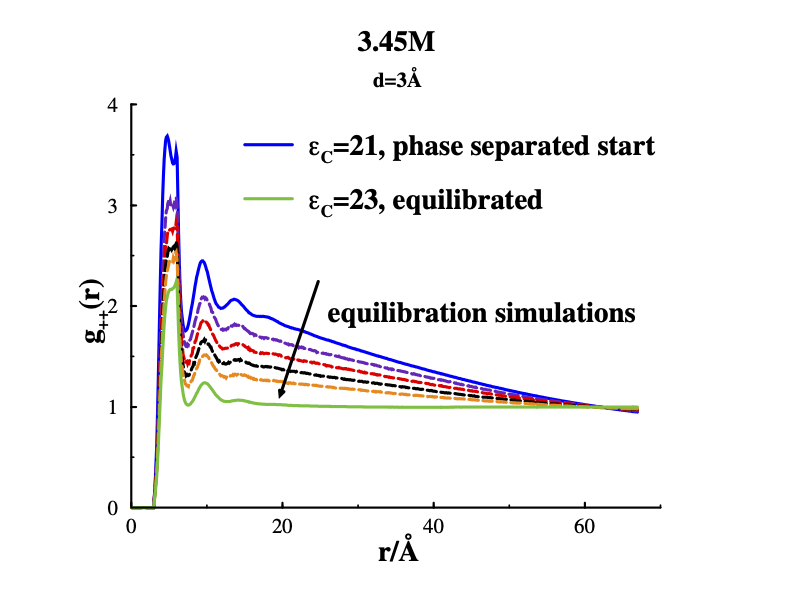}                      
    }
	\caption{(a) Effect of changing the $\varepsilon_c$ on the phase stability at 3.45 M. All systems contain 5000 ion pairs. (b) An
          illustration of the progression from a phase separated to homogenised system, for
		a system with $d=3${ \AA} and $\varepsilon_c = 23$. The phase separated system was created from
		simulations with $\varepsilon_c = 21$. Displayed are the cation-cation radial distribution functions from a set of short
                simulations (dashed lines), along with a sufficiently long  final equilibrium simulation (with $\varepsilon_c = 23$).}
	\label{fig:gpp}
\end{figure}

There are two free parameters in our model potential, which control the magnitude of the 
short-ranged attraction.  We found that at a high coupling strength, (small $d$ and $\varepsilon_c$) the
bulk solution appears to separate into an amorphous condensed phase in equilibrium with a dilute clustered phase.  This
could be viewed as the {\em saturation limit} for the solution, except in real electrolytes the 
condensed phase would be an ordered crystal.   This phase instability is seen in Figure \ref{fig:gpp}(a) where we observe the 
response of the cation-cation pair correlation $g_{++}(r)$ to changes of $\varepsilon_c$, at a concentration of 3.45 M. 
The development of a pronounced long-ranged slope in the tail of $g_{++}(r)$ is a signature of
a phase separation. This was further supported by configurational snapshots in the SI, whereby visual
inspection of the condensed phase suggests that it is non-crystalline in nature. 
While a more thorough structural analysis is lacking, we assert that determining the precise nature
of the condensed phase is anyway not pertinent to the aims of this study.   
We can, however, plausibly argue that an amorphous condensed phase is not surprising, 
given the approximate nature of our model, wherein significant many-body effects are ignored.  In particular, 
many-body effects are expected to promote clusters with ordered cores but labile outer regions, as 
dielectric saturation is larger towards the centre of the clusters, while closer to the extremities solvent
polarisation remains large.  On the other hand, the pair-wise approximation, used 
in this study,  somewhat erroneously strengthens ion-ion interactions {\em throughout} the cluster, 
as ameliorated by the choice of $\varepsilon_c$.  Thus, in order to predict a reasonable saturation 
concentration for the electrolyte model, a sensible choice for $\varepsilon_c$
should fall between the bulk value of the solvent and the (small) internal dielectric constant of a crystal.  
However, such a choice will likely stabilise an amorphous condensed phase rather than a crystalline phase, 
given that $\varepsilon_c$ will be larger than the expected dielectric response inside a crystal.  
Furthermore, the amorphous phase will also have less binding energy than 
that of the crystal, which suggests that our pair-wise approximation will possibly predict less clustering 
as the solution concentration approaches saturation.  KCl and NaCl have saturation concentrations
of about 4.5-6 M, which suggests we should consider a similar saturation concentration in our
modelling.

Given that the aim of our study was primarily to explore the impact of clustering on electrostatic correlations, 
it was important that we select a value for  $\varepsilon_c$ so that the system displayed large but finite clusters.
From Figures \ref{fig:gpp}(a)-(b)  we see that, at 3.45 M, the system phase separates for $\varepsilon_c$ 
values of 20 and 21, but not when $\varepsilon_c$ is above 22 (for $d=$ 3 Å) . Thus we have set $\varepsilon_c = 23$
for all subsequent simulations.  This value provides a significant degree of clustering in our model 
while the solution remains unsaturated at least up to 3.45 M. 
A different choice of $d$ would lead to a different choice of $\varepsilon_c$, 
as described in the Supporting Information, SI

It is of course possible that the system is {\em metastable} rather than stable for $\varepsilon_c = 23$ at 3.45 M.
However, we have made tests ensuring that a metastable scenario is highly unlikely. Specifically, we have (several times)
initiated simulations from a phase separated system, at $\varepsilon_c = 21$, only to find that, upon switching to $\varepsilon_c = 23$, 
the system transitions to a single phase state at equilibrium.
This is illustrated in Figure \ref{fig:gpp}(b), where we depict how the pronounced long-ranged gradient
of the initial radial distribution function (indicating a phase-separated system), gradually
disappears. 

\begin{figure}
	\centering
	\includegraphics{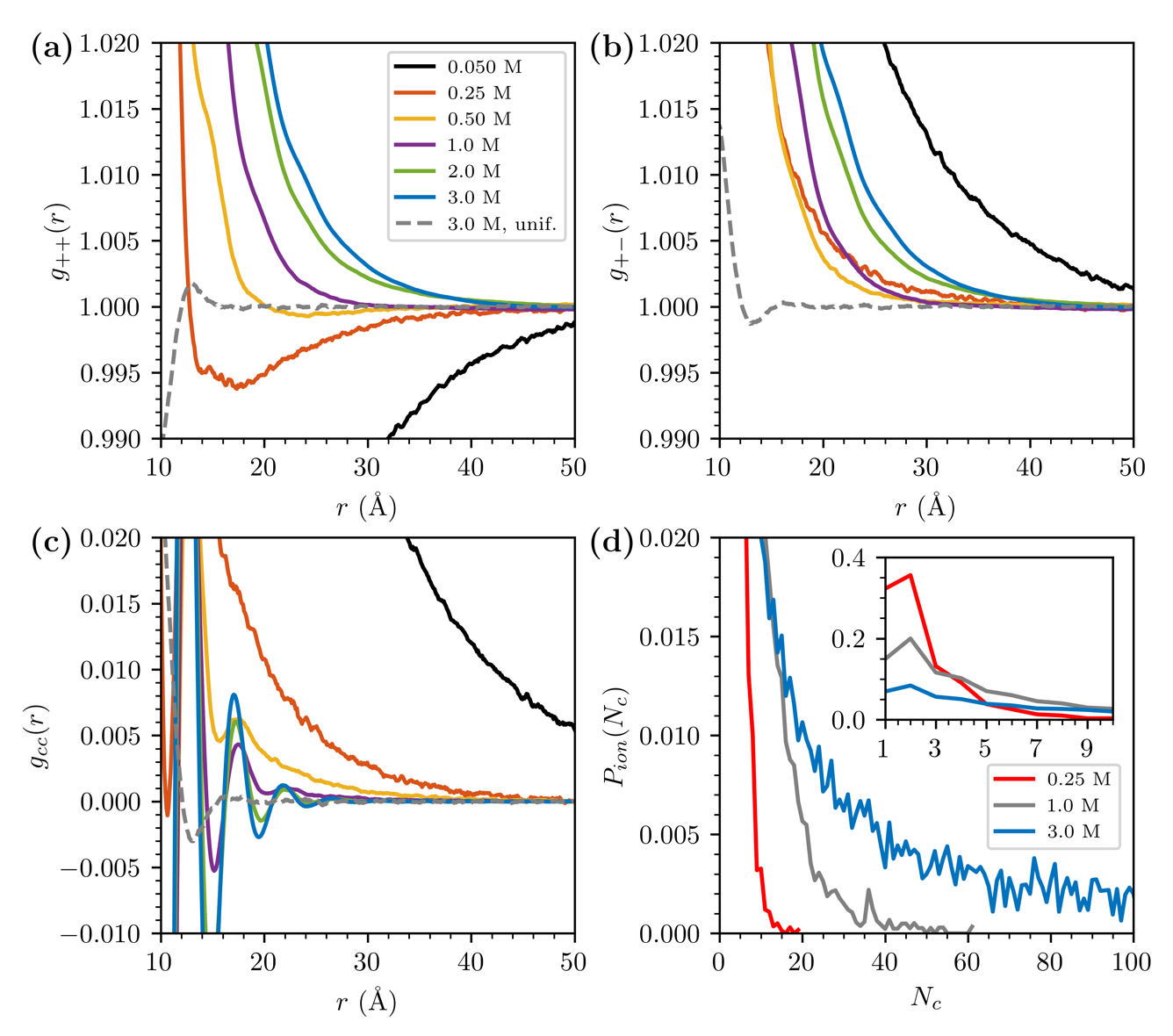}        
	\caption{Results from bulk simulations. (a) Cation-cation (or anion-anion) resolved radial distribution
          functions, $g_{++}(r)$ (on average identical to $g_{--}(r)$), for the local dielectric saturation
          model (full line), and a uniform low-dielectric constant (a uniform value of $\varepsilon_r=23$) model (dashed line).
          (In order to improve statistics we have in reality measured ``$g_{++}(r)$'' as $(g_{++}(r)+g_{--}(r))/2$)
          (b) Cation-anion resolved radial distribution functions, $g_{+-}(r)$.
          (c) $g_{cc}(r) \equiv g_{+-}(r)-(g_{++}(r)+g_{--}(r))/2$.
          (d) Probability of an ion to be part of a cluster of size $N_c$, $P_{ion}(N_c)$, at three different concentrations. An ion must
          be within a distance $\delta$ or less, of at least one other ion within a cluster, in order to be a member of that cluster.
          We have set $\delta = d+0.5${\AA}. The inset is a focus on small clusters.}
	\label{fig1}
\end{figure}

We reiterate that our potential model introduces an additional short-ranged asymmetric interaction, which
promotes association between unlike ions and the formation of large clusters. 
This cluster formation as a function of the electrolyte concentration can be readily 
observed in the species resolved radial distribution functions, $g_{ij}(r)$ (with $i,j = \pm$)
as illustrated in Figure \ref{fig1}(a)-(b).   By symmetry $g_{--}(r) = g_{++}(r)$ and $g_{+-}(r) = g_{-+}(r)$
thus it is prudent to use $(g_{++}(r)+ g_{--}(r))/2$ and $(g_{+-}(r)+ g_{-+}(r))/2$ for the like and
unlike correlation functions.  We can define the total density correlation function 
$g_{nn}(r) = (g_{++}(r)+g_{--}(r)+g_{+-}(r)+g_{-+}(r))/4$ (the particle density correlation around any ion)
as well as the so-called charge-charge correlation
function $g_{cc}(r) = (g_{+-}(r)+g_{-+}(r)-g_{++}(r)-g_{--}(r))/2$ (the counter-charge 
density around an ion).  The charge-charge correlation functions are shown in Figure \ref{fig1}(c). 
Using linear response theory it is possible to relate the interaction free energy between two charged surfaces
(at large separation) to these correlation functions at long-range.  As charged surfaces perturb both the
charge and particle densities of the electrolyte contained between them, the interaction between the 
surfaces at large separations is dictated either by $g_{cc}(r)$ or $g_{nn}(r)$, whichever has the longest range.
Classical mean-field theory, wherein ions are assumed to respond only to the mean electrostatic potential, 
asserts that $h_{cc}(r)$ has a Yukawa form,  $h_{cc}(r) \sim \exp(-r/\lambda_D)/r$ and $h_{nn}(r)$ is
much shorter-ranged  ($h_{\alpha\beta}(r) = g(r)_{\alpha\beta} - 1$).
In fact, linearised mean-field analysis predicts that the charged surfaces
will not alter the total ionic density between them at all, $h_{nn}(r) = 0$, whereas to second order 
we have $h_{nn}(r) \sim \exp(-2r/\lambda_D)/r^2$.   This qualitative picture may 
change in the presence of strong correlations and, as in our case, with the formation of large clusters.   

At the lowest concentration investigated (0.05 M), we observe what 
could be described as {\em mean-field} behaviour for the correlation functions.  
Here $g_{++}(r)$ displays a co-ion exclusion region, while $g_{+-}(r)$
illustrates counter-ion attraction, so that  $h_{++}(r) \approx -h_{+-}(r)$ 
as they approach zero. Between 0.25-0.5 M a peak appears in  
$g_{++}(r)$ at short-range, which then displays co-ion exclusion at larger distances.
As the concentration increases however, the co-ion exclusion region diminishes in size so that at
0.5 M it has almost vanished.  On the other hand, $g_{+-}(r)$ still displays an attraction between counterions, 
but its range appears to decrease, which is actually qualitatively consistent with the mean-field behaviour.  
In Figure \ref{fig1}(c), we observe the a reduction in range of  $g_{cc}(r)$ as well.
In addition, oscillations appear at short range at 0.5 M, which, together with the behaviour of  $g_{++}(r)$, suggests accumulation of ions 
into clusters.   However, even in this concentration range, the {\em tails} of these correlation
functions are still Yukawa like and as we show below (and in the SI) asymptotic fitting
gives a correlation length which is greater than $\lambda_D$, indicating
some degree of underscreening.  

At even higher concentrations (1 M and 3 M), we note the complete disappearance of the 
co-ion exclusion region in $g_{++}(r)$.  This is accompanied by apparent coincidence 
of $g_{++}(r)$ and $g_{+-}(r)$ at longer range, as indicated by the diminished range of 
their difference, $g_{cc}(r)$, as shown in Figure \ref{fig1}(c). That is, at lower 
concentrations, the long-range decays of all these correlation functions were similar,
albeit $g_{++}(r)$ and $g_{+-}(r)$ approach unity from below and above respectively.
Indeed, in the SI we show that they can be all reasonably well fitted to a Yukawa form,
$\sim \exp(-r/\lambda)/r$, where $\lambda$ is generally larger than the Debye length.  
However, at 1 M and 3 M, $g_{++}(r)$ and  $g_{+-}(r)$ both approach unity from
above and, their range is significantly longer than $g_{cc}(r)$.  These results indicate
that there is a great deal of clustering occurring, causing much longer range 
correlations in the density, $g_{nn}(r) $, compared with 
the charge, $g_{cc}(r)$.  The correlation length of  $g_{nn}(r)$ will be determined by the 
typical cluster size.  In Figure \ref{fig1}(d), we show 
the probability, $P_{ion}(N_c)$, that an ion is a member of a cluster of size $N_c$
for different concentrations, and the growth in cluster sizes is apparent
as the ion concentration increases.  The growth in clusters is accompanied
with more rapid screening of charges.  This effect on charge screening is 
also qualitatively predicted by mean-field  theories, as the Debye length 
decreases with concentration.  But here we also see that $g_{cc}(r)$ loses its 
Yukawa form and becomes oscillatory, so hard-core correlations within 
dense clusters are clearly playing a role. Most clusters are expected to be close to neutral, 
which is why the difference between cation-cation and cation-anion correlations at long range 
vanishes.  What has not been previously reported in 
other theoretical treatments of electrolyte models (as far as we are aware),
is the significant {\em increase} in range of  $g_{nn}(r)$ as a function of
concentration, compared with $g_{cc}(r)$.  

Thus, our model predicts that at around 1 M concentration, the interaction between charged
surfaces will begin to be dominated by density correlations, as measured by
$g_{nn}(r)$, rather than charge correlations (from  $g_{cc}(r)$), which 
were significant at lower concentrations.  In our model, density correlations are 
affected by cluster formation and their typical size, whereas 
charge correlations are determined by the availability of screening charges.

We argue that our modification of the RPM gives rise to clusters that are much
larger than usual applications of the RPM.  
To illustrate this, we also calculated the correlation functions at 3 M for a simple RPM case, where the uniform
dielectric constant was chosen to be $\varepsilon_r=\varepsilon_c = 23$.  That is, we assumed 
dielectric saturation occurs over the {\em full range} of the Coulomb interaction, rather
than just close to ion-ion contact.  We
see (Figure \ref{fig1}(a)) that $g_{++}(r)$ then displays a short-ranged
co-ion exclusion region superimposed on an oscillatory profile, with no long-ranged
tail as observed in our modified RPM.  Similarly,  $g_{+-}(r)$ shows a short-ranged
counter-ion enhancement with oscillations, again without a long-ranged tail (Figure \ref{fig1}(b) .  
The corresponding $g_{cc}(r)$ is also oscillatory, but with amplitude much smaller than the 
modified RPM (Figure \ref{fig1}(c)), which is evidence of much larger clustering induced by the 
modified RPM.  It is clear that the additional short-ranged interaction introduced in our
modified RPM reduces the incentive for ions to dissociate from clusters once they are formed.  
If dielectric saturation is assumed to occur everywhere, even in regions of low ionic density,
there is less of an advantage for ions to cluster.  Note, that this affect is ameliorated by the
decreased in ion solvation expected to occur in clusters, an effect which must 
be accounted for when choosing a suitable value for $\varepsilon_c$.

It may seem surprising that the {\em range} of a reduced dielectric response has such a strong
impact on the cluster forming tendency. However, one should take cognisance of the fact that
with a uniform and low dielectric constant, there are inevitable strong
repulsions between like charges in a cluster. These repulsions are considerably weaker if
the reduced dielectric response is local, mainly  influencing the interaction
between charges of opposite sign.

\begin{figure} 
	\centering
	\includegraphics{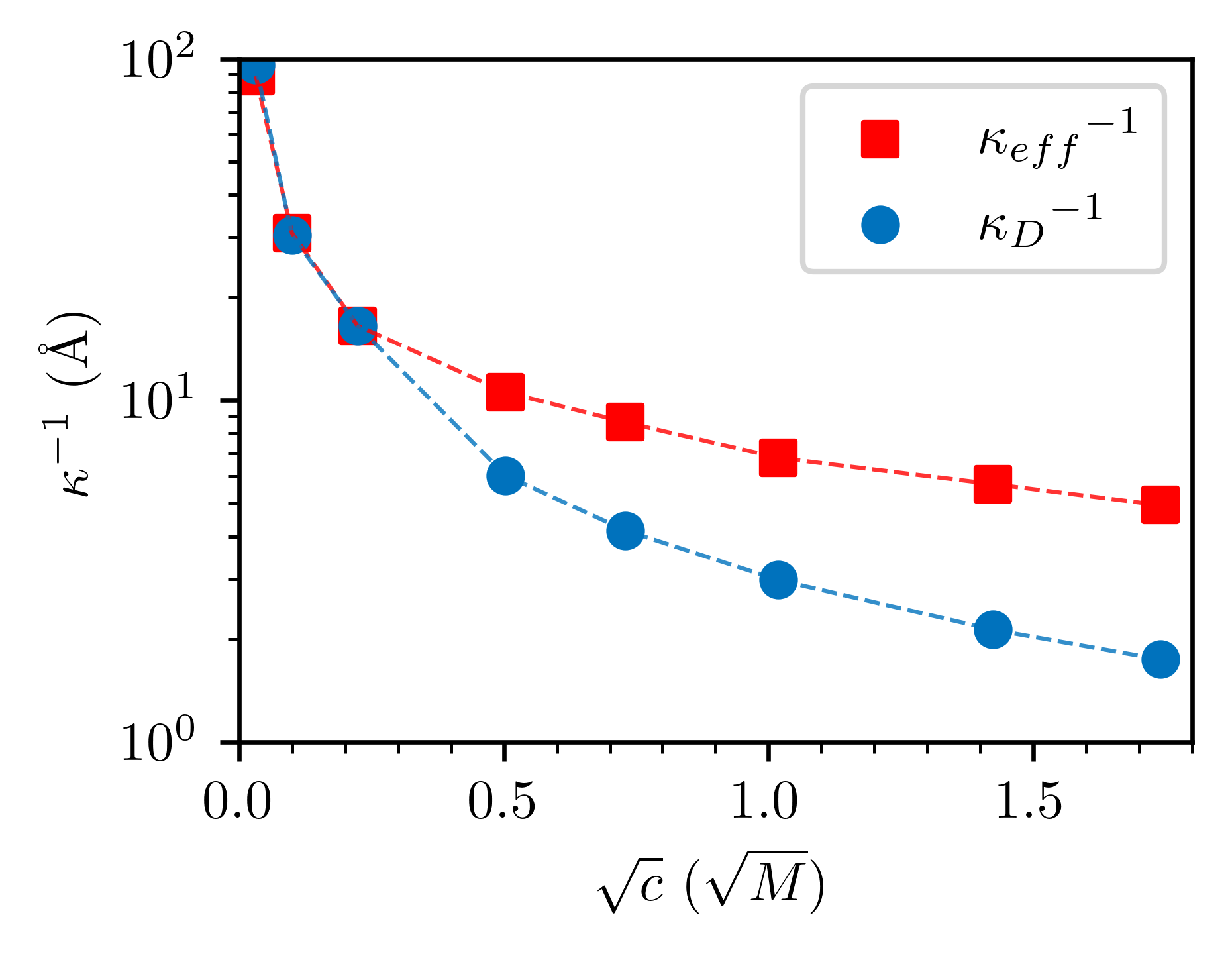}        
	\caption{Effective electrostatic screening lengths obtained via the modified Widom
          technique, and the corresponding Debye screening
          lengths. The dashed lines guide the eye.}
	\label{fig2}
\end{figure} 

In previous work, we developed a simulation method that uses a modified Widom technique to estimate the long range
decay length of $g_{cc}(r)$, which is very accurate if one can assume a Yukawa form,
$g_{cc}(r) \sim 1+ A\exp(-\kappa_{eff} r)/r$ \cite{Forsman:24a}. The method is briefly described in the SI.
The results for the correlation functions calculated above are compared with the standard Debye screening
length ($\kappa_D^2 = \beta c \sum_i (e z_i)^2$) in Figure \ref{fig2}.  Note that while the Yukawa form is only
really valid below 1 M, our method is nevertheless still able to extract an {\em effective} screening length at the 
higher concentrations.  This is because it is based on a linear expansion of the free energy functional
that predicts a screening length from an ensemble average.  This average is still calculable even in
cases where assumptions which lead to a Yukawa form break down.  In Figure \ref{fig2}
we do observe some underscreening, the degree of which is much smaller  
than that suggested by SFA experiments\cite{Gebbie:2013, Gebbie:2015, Smith:2016, D3FD00042G, C6FD00250A}. 
In particular, we do not observe an increase in screening length with concentration.  
In any case, as described above,  the decay length of  $g_{cc}(r)$ is not the dominant 
one above 1 M in any case, but rather that of  $g_{nn}(r)$.
Interestingly, by the use of asymptotic analysis techniques, researchers have previously
determined that such a scenario 
is theoretically possible \cite{LeotedeCarvalho1994, Attard_1993}.   That is, even
in the RPM, a transition may occur from charge-charge correlation to a density-density dominated correlation 
in the asymptotic (long-range) regime.  
\begin{figure}
	\centering
	\includegraphics{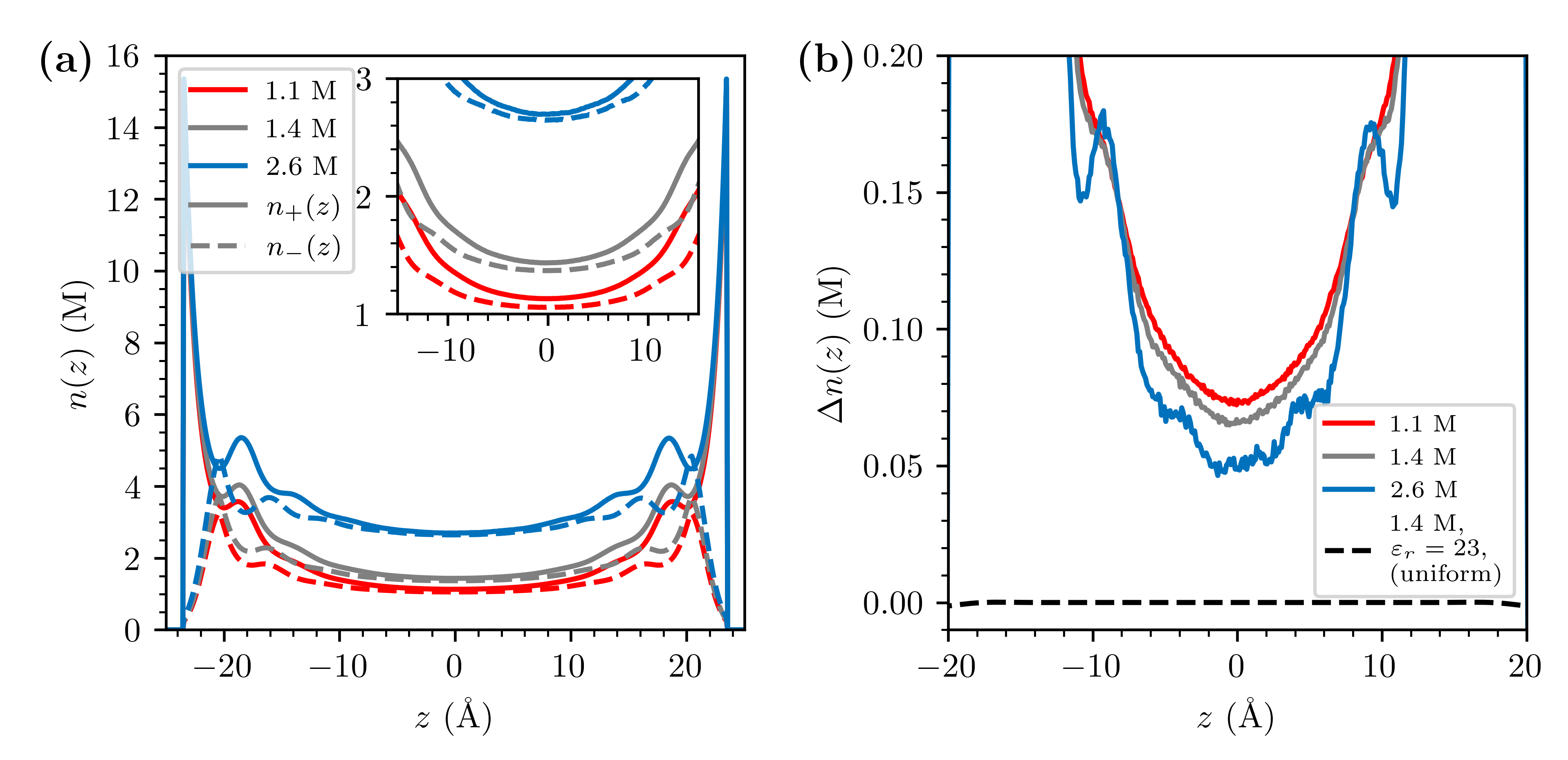}                
	\caption{Results from the slit geometry simulations, with negatively charged surfaces. (a) Concentrations of
          cations (solid lines) and anions (dashed lines) along
          the $z$-axis, for three different simulated bulk
          concentrations (these are estimated from mid plane values).
          (b) The concentration difference between cations and anions along the $z$-axis.
        Here, we have also added data from RPM simulations with a uniform dielectric constant of 23, at 1.4M (dashed line).}
	\label{fig:walls}
\end{figure}

Finally, we turn our attention to structural properties of our modified RPM, 
in the presence of two macroscopic flat and negatively charged surfaces.
For these systems, we have chosen a simulation model with a slit geometry. 
The slit simulations enable us to study the behaviour of the model in a system 
that approximates the experimental SFA setup: an electrolyte confined between
two macroscopic charged surfaces. We investigated three concentrations at
a constant inverse surface charge density of $-70\,e/${\AA}$^2$.  
The resulting ion density distributions, $n_+(z)$ and
$n_-(z)$, where $z$ is the direction normal to the surfaces, are presented in
Figure \ref{fig:walls}(a). We can observe a significant difference between $n_+(z)$ and $n_-(z)$ 
at the mid-plane between the surfaces, even when these are
50 {\AA} apart. This is in stark contrast to mean-field predictions, since the mid plane 
is more than 13 Debye lengths distant from the surfaces, at the highest
investigated bulk concentration (about 2.6 M). In order to clarify our results this further, 
we plot the density difference, $\Delta n(z) \equiv n_+(z) - n_-(z)$, in
Figure \ref{fig:walls}(b). Even though the overall $\Delta n(z)$ profile does drop 
as the salt concentration increases, we note that the salt dependence is quite weak, 
and that a significant long-ranged tail persists also for very high ionic strengths.
The scenario would again be quite different with a model using a uniform 
dielectric constant (even $\varepsilon_r$=23), in which case $\Delta n(z)$ rapidly vanishes
away from a charged surface, at high concentrations. This is explicitly shown by 
the dashed grey line in Figure \ref{fig:walls}(b). While
we have not measured the net interaction between such charged surfaces in this study, 
the long-ranged tail of $\Delta n(z)$ clearly implies a slowly decaying
surface force. This force will be quantified in future simulation work.

Considering the fact that experimental approaches extract the effective 
correlation lengths\cite{C6FD00250A}, which may or may not be purely a result
of charge-charge correlations, the observation of a transition from the
expected charge-charge asymptotic domination to a density-density domination 
at high concentrations strongly supports the hypothesis that
cluster formation, has a crucial influence on anomalous underscreening effects.

This theoretical study has investigated the influence of local dielectric saturation on the structure of electrolytes modelled by the RPM. 
We demonstrate that local dielectric saturation induces significant
clustering. Interactions between such clusters dominate the asymptotic correlations for systems approaching or exceeding concentrations
of about 1 M. When charged surfaces are immersed in such solutions, a net charge density develops, that decays quite slowly with the
transverse distance to these surfaces. This suggests that there might be long-ranged interactions between such surfaces, commensurate
with observations by SFA.

\begin{acknowledgement}
  J.F. acknowledges financial support by the Swedish Research Council, and computational resources by the
  Lund University computer cluster organisation, LUNARC.

\end{acknowledgement}

\begin{suppinfo}

The following files are available free of charge.
\begin{itemize}
  \item Supporting information: Detailed simulation methods, and further analyses.
  \item Github repository: all codes used for simulations, along with the data generated, is freely available.
\end{itemize}

\end{suppinfo}


\begin{mcitethebibliography}{41}
\providecommand*\natexlab[1]{#1}
\providecommand*\mciteSetBstSublistMode[1]{}
\providecommand*\mciteSetBstMaxWidthForm[2]{}
\providecommand*\mciteBstWouldAddEndPuncttrue
  {\def\EndOfBibitem{\unskip.}}
\providecommand*\mciteBstWouldAddEndPunctfalse
  {\let\EndOfBibitem\relax}
\providecommand*\mciteSetBstMidEndSepPunct[3]{}
\providecommand*\mciteSetBstSublistLabelBeginEnd[3]{}
\providecommand*\EndOfBibitem{}
\mciteSetBstSublistMode{f}
\mciteSetBstMaxWidthForm{subitem}{(\alph{mcitesubitemcount})}
\mciteSetBstSublistLabelBeginEnd
  {\mcitemaxwidthsubitemform\space}
  {\relax}
  {\relax}

\bibitem[Israelachvili(1991)]{Israelachvili:91}
Israelachvili,~J.~N. \emph{Intermolecular and Surface Forces, 2nd Ed.};
  Academic Press: London, 1991\relax
\mciteBstWouldAddEndPuncttrue
\mciteSetBstMidEndSepPunct{\mcitedefaultmidpunct}
{\mcitedefaultendpunct}{\mcitedefaultseppunct}\relax
\EndOfBibitem
\bibitem[Evans and Wennerstr{\"o}m(1994)Evans, and Wennerstr{\"o}m]{Evans:94}
Evans,~F.~A.; Wennerstr{\"o}m,~H. \emph{The colloidal domain: where Physics,
  Chemistry, Biology and Technology meet}; VCH Publishers: New York, 1994\relax
\mciteBstWouldAddEndPuncttrue
\mciteSetBstMidEndSepPunct{\mcitedefaultmidpunct}
{\mcitedefaultendpunct}{\mcitedefaultseppunct}\relax
\EndOfBibitem
\bibitem[Holm \latin{et~al.}(2001)Holm, Kekicheff, and Podgornik]{Holm:01}
Holm,~C.; Kekicheff,~P.; Podgornik,~R. \emph{Electrostatic Effects in Soft
  Matter and Biophysics}; Kluwer Academic Publishers: Dordrecht, 2001\relax
\mciteBstWouldAddEndPuncttrue
\mciteSetBstMidEndSepPunct{\mcitedefaultmidpunct}
{\mcitedefaultendpunct}{\mcitedefaultseppunct}\relax
\EndOfBibitem
\bibitem[Derjaguin and Landau(1941)Derjaguin, and Landau]{Derjaguin:41}
Derjaguin,~B.~V.; Landau,~L. Theory of the Stability of Strongly Charged
  Lyophobic Sols and of the Adhesion of Strongly Charged Particles in Solutions
  of Electrolytes. \emph{Acta Phys. Chim. URSS} \textbf{1941}, \emph{14},
  633--662\relax
\mciteBstWouldAddEndPuncttrue
\mciteSetBstMidEndSepPunct{\mcitedefaultmidpunct}
{\mcitedefaultendpunct}{\mcitedefaultseppunct}\relax
\EndOfBibitem
\bibitem[Verwey and Overbeek(1948)Verwey, and Overbeek]{Verwey:48}
Verwey,~E. J.~W.; Overbeek,~J. T.~G. \emph{Theory of the Stability of Lyophobic
  Colloids}; Elsevier Publishing Company Inc.: Amsterdam, 1948\relax
\mciteBstWouldAddEndPuncttrue
\mciteSetBstMidEndSepPunct{\mcitedefaultmidpunct}
{\mcitedefaultendpunct}{\mcitedefaultseppunct}\relax
\EndOfBibitem
\bibitem[Nordholm(1984)]{Nordholm:84a}
Nordholm,~S. Generalized van der Waals theory. XII. Application to ionic
  solutions. \emph{Aust. J. Chem.} \textbf{1984}, \emph{37}, 1\relax
\mciteBstWouldAddEndPuncttrue
\mciteSetBstMidEndSepPunct{\mcitedefaultmidpunct}
{\mcitedefaultendpunct}{\mcitedefaultseppunct}\relax
\EndOfBibitem
\bibitem[Guldbrand \latin{et~al.}(1984)Guldbrand, J{\"o}nsson, Wennerstr{\"o}m,
  and Linse]{Guldbrand:84}
Guldbrand,~L.; J{\"o}nsson,~B.; Wennerstr{\"o}m,~H.; Linse,~P. Electrical
  Double Layer Forces, {A Monte Carlo} Study. \emph{J. Chem. Phys.}
  \textbf{1984}, \emph{80}, 2221\relax
\mciteBstWouldAddEndPuncttrue
\mciteSetBstMidEndSepPunct{\mcitedefaultmidpunct}
{\mcitedefaultendpunct}{\mcitedefaultseppunct}\relax
\EndOfBibitem
\bibitem[Kjellander and Marcelja(1986)Kjellander, and Marcelja]{Kjellander:86}
Kjellander,~R.; Marcelja,~S. Interaction of charged surfaces in electrolyte
  solutions. \emph{Chem. Phys. Lett.} \textbf{1986}, \emph{127}, 402--407\relax
\mciteBstWouldAddEndPuncttrue
\mciteSetBstMidEndSepPunct{\mcitedefaultmidpunct}
{\mcitedefaultendpunct}{\mcitedefaultseppunct}\relax
\EndOfBibitem
\bibitem[Valleau \latin{et~al.}(1991)Valleau, Ivkov, and Torrie]{Torrie:91}
Valleau,~J.; Ivkov,~R.; Torrie,~G.~M. \emph{J. Phys. Chem.} \textbf{1991},
  \emph{95}, 520\relax
\mciteBstWouldAddEndPuncttrue
\mciteSetBstMidEndSepPunct{\mcitedefaultmidpunct}
{\mcitedefaultendpunct}{\mcitedefaultseppunct}\relax
\EndOfBibitem
\bibitem[Gebbie \latin{et~al.}(2013)Gebbie, Valtiner, Banquy, Fox, Henderson,
  and Israelachvili]{Gebbie:2013}
Gebbie,~M.~A.; Valtiner,~M.; Banquy,~X.; Fox,~E.~T.; Henderson,~W.~A.;
  Israelachvili,~J.~N. Ionic liquids behave as dilute electrolyte solutions.
  \emph{PNAS} \textbf{2013}, \emph{110}, 9674--9679\relax
\mciteBstWouldAddEndPuncttrue
\mciteSetBstMidEndSepPunct{\mcitedefaultmidpunct}
{\mcitedefaultendpunct}{\mcitedefaultseppunct}\relax
\EndOfBibitem
\bibitem[Gebbie \latin{et~al.}(2015)Gebbie, Dobbs, Valtiner, and
  Israelachvili]{Gebbie:2015}
Gebbie,~M.~A.; Dobbs,~H.~A.; Valtiner,~M.; Israelachvili,~J.~N. Long-range
  electrostatic screening in ionic liquids. \emph{PNAS} \textbf{2015},
  \emph{112}, 7432--7437\relax
\mciteBstWouldAddEndPuncttrue
\mciteSetBstMidEndSepPunct{\mcitedefaultmidpunct}
{\mcitedefaultendpunct}{\mcitedefaultseppunct}\relax
\EndOfBibitem
\bibitem[Smith \latin{et~al.}(2016)Smith, Lee, and Perkin]{Smith:2016}
Smith,~A.~M.; Lee,~A.~A.; Perkin,~S. The Electrostatic Screening Length in
  Concentrated Electrolytes Increases with Concentration. \emph{J. Phys. Chem.
  Lett.} \textbf{2016}, \emph{7}, 2157--2163\relax
\mciteBstWouldAddEndPuncttrue
\mciteSetBstMidEndSepPunct{\mcitedefaultmidpunct}
{\mcitedefaultendpunct}{\mcitedefaultseppunct}\relax
\EndOfBibitem
\bibitem[Fung and Perkin(2023)Fung, and Perkin]{D3FD00042G}
Fung,~Y. K.~C.; Perkin,~S. Structure and anomalous underscreening in
  ethylammonium nitrate solutions confined between two mica surfaces.
  \emph{Faraday Discuss.} \textbf{2023}, \emph{246}, 370--386\relax
\mciteBstWouldAddEndPuncttrue
\mciteSetBstMidEndSepPunct{\mcitedefaultmidpunct}
{\mcitedefaultendpunct}{\mcitedefaultseppunct}\relax
\EndOfBibitem
\bibitem[Lee \latin{et~al.}(2017)Lee, Perez-Martinez, Smith, and
  Perkin]{C6FD00250A}
Lee,~A.~A.; Perez-Martinez,~C.~S.; Smith,~A.~M.; Perkin,~S. Underscreening in
  concentrated electrolytes. \emph{Faraday Discuss.} \textbf{2017}, \emph{199},
  239--259\relax
\mciteBstWouldAddEndPuncttrue
\mciteSetBstMidEndSepPunct{\mcitedefaultmidpunct}
{\mcitedefaultendpunct}{\mcitedefaultseppunct}\relax
\EndOfBibitem
\bibitem[Yuan \latin{et~al.}(2022)Yuan, Deng, Zhu, Liu, and Craig]{Yuan:22}
Yuan,~H.; Deng,~W.; Zhu,~X.; Liu,~G.; Craig,~V. S.~J. Colloidal Systems in
  Concentrated Electrolyte Solutions Exhibit Re-entrant Long-Range
  Electrostatic Interactions due to Underscreening. \emph{Langmuir}
  \textbf{2022}, \emph{38}, 6164--6173\relax
\mciteBstWouldAddEndPuncttrue
\mciteSetBstMidEndSepPunct{\mcitedefaultmidpunct}
{\mcitedefaultendpunct}{\mcitedefaultseppunct}\relax
\EndOfBibitem
\bibitem[Attard(1993)]{Attard_1993}
Attard,~P. Asymptotic analysis of primitive model electrolytes and the
  electrical double layer. \emph{Phys. Rev. E} \textbf{1993}, \emph{48},
  3604--3621\relax
\mciteBstWouldAddEndPuncttrue
\mciteSetBstMidEndSepPunct{\mcitedefaultmidpunct}
{\mcitedefaultendpunct}{\mcitedefaultseppunct}\relax
\EndOfBibitem
\bibitem[Coupette \latin{et~al.}(2018)Coupette, Lee, and
  Härtel]{coupette_screening_2018}
Coupette,~F.; Lee,~A.~A.; Härtel,~A. Screening {Lengths} in {Ionic} {Fluids}.
  \emph{Phys. Rev. Lett.} \textbf{2018}, \emph{121}, 075501\relax
\mciteBstWouldAddEndPuncttrue
\mciteSetBstMidEndSepPunct{\mcitedefaultmidpunct}
{\mcitedefaultendpunct}{\mcitedefaultseppunct}\relax
\EndOfBibitem
\bibitem[H{\"a}rtel \latin{et~al.}(2023)H{\"a}rtel, B\"ultmann, and
  Coupette]{Hartel:23}
H{\"a}rtel,~A.; B\"ultmann,~M.; Coupette,~F. Anomalous Underscreening in the
  Restricted Primitive Model. \emph{Phys. Rev. Lett.} \textbf{2023},
  \emph{130}, 108202\relax
\mciteBstWouldAddEndPuncttrue
\mciteSetBstMidEndSepPunct{\mcitedefaultmidpunct}
{\mcitedefaultendpunct}{\mcitedefaultseppunct}\relax
\EndOfBibitem
\bibitem[Kumar \latin{et~al.}(2022)Kumar, Cats, Alotaibi, Ayirala, Yousef, {van
  Roij}, Siretanu, and Mugele]{Kumar:22}
Kumar,~S.; Cats,~P.; Alotaibi,~M.~B.; Ayirala,~S.~C.; Yousef,~A.~A.; {van
  Roij},~R.; Siretanu,~I.; Mugele,~F. Absence of anomalous underscreening in
  highly concentrated aqueous electrolytes confined between smooth silica
  surfaces. \emph{J. Colloid Interface Sci.} \textbf{2022}, \emph{622},
  819--827\relax
\mciteBstWouldAddEndPuncttrue
\mciteSetBstMidEndSepPunct{\mcitedefaultmidpunct}
{\mcitedefaultendpunct}{\mcitedefaultseppunct}\relax
\EndOfBibitem
\bibitem[Rotenberg \latin{et~al.}(2018)Rotenberg, Bernard, and
  Hansen]{Rotenberg:18}
Rotenberg,~B.; Bernard,~O.; Hansen,~J.-P. Underscreening in ionic liquids: a
  first principles analysis. \emph{J. Phys. Condens. Matter} \textbf{2018},
  \emph{30}, 054005\relax
\mciteBstWouldAddEndPuncttrue
\mciteSetBstMidEndSepPunct{\mcitedefaultmidpunct}
{\mcitedefaultendpunct}{\mcitedefaultseppunct}\relax
\EndOfBibitem
\bibitem[Kjellander(2020)]{Kjellander:20}
Kjellander,~R. A multiple decay-length extension of the Debye-Hückel theory:
  to achieve high accuracy also for concentrated solutions and explain
  under-screening in dilute symmetric electrolytes. \emph{Phys. Chem. Chem.
  Phys.} \textbf{2020}, \emph{22}, 23952--23985\relax
\mciteBstWouldAddEndPuncttrue
\mciteSetBstMidEndSepPunct{\mcitedefaultmidpunct}
{\mcitedefaultendpunct}{\mcitedefaultseppunct}\relax
\EndOfBibitem
\bibitem[Coles \latin{et~al.}(2020)Coles, Park, Nikam, Kanduc, Dzubiella, and
  Rotenberg]{Coles:20}
Coles,~S.~W.; Park,~C.; Nikam,~R.; Kanduc,~M.; Dzubiella,~J.; Rotenberg,~B.
  Correlation Length in Concentrated Electrolytes: Insights from All-Atom
  Molecular Dynamics Simulations. \emph{J. Phys. Chem. B} \textbf{2020},
  \emph{124}, 1778--1786\relax
\mciteBstWouldAddEndPuncttrue
\mciteSetBstMidEndSepPunct{\mcitedefaultmidpunct}
{\mcitedefaultendpunct}{\mcitedefaultseppunct}\relax
\EndOfBibitem
\bibitem[Zeman \latin{et~al.}(2020)Zeman, Kondrat, and Holm]{Zeman:20}
Zeman,~J.; Kondrat,~S.; Holm,~C. Bulk ionic screening lengths from extremely
  large-scale molecular dynamics simulations. \emph{Chem. Commun.}
  \textbf{2020}, \emph{56}, 15635--15638\relax
\mciteBstWouldAddEndPuncttrue
\mciteSetBstMidEndSepPunct{\mcitedefaultmidpunct}
{\mcitedefaultendpunct}{\mcitedefaultseppunct}\relax
\EndOfBibitem
\bibitem[Cats \latin{et~al.}(2021)Cats, Evans, H{\"a}rtel, and van
  Roij]{Cats:21}
Cats,~P.; Evans,~R.; H{\"a}rtel,~A.; van Roij,~R. {Primitive model electrolytes
  in the near and far field: Decay lengths from DFT and simulations}. \emph{J.
  Chem. Phys.} \textbf{2021}, \emph{154}, 124504\relax
\mciteBstWouldAddEndPuncttrue
\mciteSetBstMidEndSepPunct{\mcitedefaultmidpunct}
{\mcitedefaultendpunct}{\mcitedefaultseppunct}\relax
\EndOfBibitem
\bibitem[Ma \latin{et~al.}(2015)Ma, Forsman, and Woodward]{Ma:2015}
Ma,~K.; Forsman,~J.; Woodward,~C.~E. Influence of ion pairing in ionic liquids
  on electrical double layer structures and surface force using classical
  density functional approach. \emph{The Journal of Chemical Physics}
  \textbf{2015}, \emph{142}, 174704\relax
\mciteBstWouldAddEndPuncttrue
\mciteSetBstMidEndSepPunct{\mcitedefaultmidpunct}
{\mcitedefaultendpunct}{\mcitedefaultseppunct}\relax
\EndOfBibitem
\bibitem[Hasted \latin{et~al.}(2004)Hasted, Ritson, and
  Collie]{hasted_dielectric_2004}
Hasted,~J.~B.; Ritson,~D.~M.; Collie,~C.~H. Dielectric {Properties} of
  {Aqueous} {Ionic} {Solutions}. {Parts} {I} and {II}. \emph{J. Chem. Phys.}
  \textbf{2004}, \emph{16}, 1--21\relax
\mciteBstWouldAddEndPuncttrue
\mciteSetBstMidEndSepPunct{\mcitedefaultmidpunct}
{\mcitedefaultendpunct}{\mcitedefaultseppunct}\relax
\EndOfBibitem
\bibitem[de~Souza \latin{et~al.}(2022)de~Souza, Kornyshev, and
  Bazant]{deSouza2022}
de~Souza,~J.; Kornyshev,~A.~A.; Bazant,~M.~Z. Polar liquids at charged
  interfaces: A dipolar shell theory. \emph{J. Chem. Phys.} \textbf{2022},
  \emph{156}\relax
\mciteBstWouldAddEndPuncttrue
\mciteSetBstMidEndSepPunct{\mcitedefaultmidpunct}
{\mcitedefaultendpunct}{\mcitedefaultseppunct}\relax
\EndOfBibitem
\bibitem[Conway and Marshall(1983)Conway, and Marshall]{Conway83}
Conway,~B.; Marshall,~S. Some common problems concerning solvent polarization
  and dielectric behaviour at ions and electrode interfaces. \emph{Aust. J.
  Chem.} \textbf{1983}, \emph{36}, 2145--2161\relax
\mciteBstWouldAddEndPuncttrue
\mciteSetBstMidEndSepPunct{\mcitedefaultmidpunct}
{\mcitedefaultendpunct}{\mcitedefaultseppunct}\relax
\EndOfBibitem
\bibitem[Bonthuis \latin{et~al.}(2012)Bonthuis, Gekle, and Netz]{Bonthuis2012}
Bonthuis,~D.~J.; Gekle,~S.; Netz,~R.~R. Profile of the Static Permittivity
  Tensor of Water at Interfaces: Consequences for Capacitance, Hydration
  Interaction and Ion Adsorption. \emph{Langmuir} \textbf{2012}, \emph{28},
  7679–7694\relax
\mciteBstWouldAddEndPuncttrue
\mciteSetBstMidEndSepPunct{\mcitedefaultmidpunct}
{\mcitedefaultendpunct}{\mcitedefaultseppunct}\relax
\EndOfBibitem
\bibitem[Danielewicz-Ferchmin \latin{et~al.}(2013)Danielewicz-Ferchmin,
  Banachowicz, and Ferchmin]{DanielewiczFerchmin2013}
Danielewicz-Ferchmin,~I.; Banachowicz,~E.; Ferchmin,~A. Dielectric saturation
  in water as quantitative measure of formation of well-defined hydration
  shells of ions at various temperatures and pressures. Vapor–liquid
  equilibrium case. \emph{J. Mol. Liq.} \textbf{2013}, \emph{187},
  157–164\relax
\mciteBstWouldAddEndPuncttrue
\mciteSetBstMidEndSepPunct{\mcitedefaultmidpunct}
{\mcitedefaultendpunct}{\mcitedefaultseppunct}\relax
\EndOfBibitem
\bibitem[Adar \latin{et~al.}(2018)Adar, Markovich, Levy, Orland, and
  Andelman]{adar_dielectric_2018}
Adar,~R.~M.; Markovich,~T.; Levy,~A.; Orland,~H.; Andelman,~D. Dielectric
  constant of ionic solutions: {Combined} effects of correlations and excluded
  volume. \emph{J. Chem. Phys.} \textbf{2018}, \emph{149}, 054504\relax
\mciteBstWouldAddEndPuncttrue
\mciteSetBstMidEndSepPunct{\mcitedefaultmidpunct}
{\mcitedefaultendpunct}{\mcitedefaultseppunct}\relax
\EndOfBibitem
\bibitem[Underwood and Bourg(2022)Underwood, and Bourg]{Underwood2022}
Underwood,~T.~R.; Bourg,~I.~C. Dielectric Properties of Water in Charged
  Nanopores. \emph{The Journal of Physical Chemistry B} \textbf{2022},
  \emph{126}, 2688--2698, PMID: 35362980\relax
\mciteBstWouldAddEndPuncttrue
\mciteSetBstMidEndSepPunct{\mcitedefaultmidpunct}
{\mcitedefaultendpunct}{\mcitedefaultseppunct}\relax
\EndOfBibitem
\bibitem[Ben-Yaakov \latin{et~al.}(2011)Ben-Yaakov, Andelman, and
  Podgornik]{ben-yaakov_dielectric_2011}
Ben-Yaakov,~D.; Andelman,~D.; Podgornik,~R. Dielectric decrement as a source of
  ion-specific effects. \emph{J. Chem. Phys.} \textbf{2011}, \emph{134},
  074705\relax
\mciteBstWouldAddEndPuncttrue
\mciteSetBstMidEndSepPunct{\mcitedefaultmidpunct}
{\mcitedefaultendpunct}{\mcitedefaultseppunct}\relax
\EndOfBibitem
\bibitem[Hubbard \latin{et~al.}(1979)Hubbard, Colonomos, and
  Wolynes]{hubbard_molecular_1979}
Hubbard,~J.~B.; Colonomos,~P.; Wolynes,~P.~G. Molecular theory of solvated ion
  dynamics. {III}. {The} kinetic dielectric decrement. \emph{J. Chem. Phys.}
  \textbf{1979}, \emph{71}, 2652--2661\relax
\mciteBstWouldAddEndPuncttrue
\mciteSetBstMidEndSepPunct{\mcitedefaultmidpunct}
{\mcitedefaultendpunct}{\mcitedefaultseppunct}\relax
\EndOfBibitem
\bibitem[Knight and Hub(2015)Knight, and Hub]{Knight2015}
Knight,~C.~J.; Hub,~J.~S. WAXSiS: a web server for the calculation of SAXS/WAXS
  curves based on explicit-solvent molecular dynamics. \emph{Nucleic Acids
  Research} \textbf{2015}, \emph{43}, W225–W230\relax
\mciteBstWouldAddEndPuncttrue
\mciteSetBstMidEndSepPunct{\mcitedefaultmidpunct}
{\mcitedefaultendpunct}{\mcitedefaultseppunct}\relax
\EndOfBibitem
\bibitem[Hansen \latin{et~al.}(2021)Hansen, Uthayakumar, Pedersen, Egelhaaf,
  and Platten]{Hansen2021}
Hansen,~J.; Uthayakumar,~R.; Pedersen,~J.~S.; Egelhaaf,~S.~U.; Platten,~F.
  Interactions in protein solutions close to liquid–liquid phase separation:
  ethanol reduces attractions via changes of the dielectric solution
  properties. \emph{Physical Chemistry Chemical Physics} \textbf{2021},
  \emph{23}, 22384–22394\relax
\mciteBstWouldAddEndPuncttrue
\mciteSetBstMidEndSepPunct{\mcitedefaultmidpunct}
{\mcitedefaultendpunct}{\mcitedefaultseppunct}\relax
\EndOfBibitem
\bibitem[Hansen \latin{et~al.}(2022)Hansen, Pedersen, Pedersen, Egelhaaf, and
  Platten]{Hansen2022}
Hansen,~J.; Pedersen,~J.~N.; Pedersen,~J.~S.; Egelhaaf,~S.~U.; Platten,~F.
  Universal effective interactions of globular proteins close to
  liquid–liquid phase separation: Corresponding-states behavior reflected in
  the structure factor. \emph{The Journal of Chemical Physics} \textbf{2022},
  \emph{156}\relax
\mciteBstWouldAddEndPuncttrue
\mciteSetBstMidEndSepPunct{\mcitedefaultmidpunct}
{\mcitedefaultendpunct}{\mcitedefaultseppunct}\relax
\EndOfBibitem
\bibitem[Forsman \latin{et~al.}()Forsman, Ribar, and Woodward]{Forsman:24a}
Forsman,~J.; Ribar,~D.; Woodward,~C.~E. \emph{submitted to Physical Chemistry
  Chemical Physics} \relax
\mciteBstWouldAddEndPunctfalse
\mciteSetBstMidEndSepPunct{\mcitedefaultmidpunct}
{}{\mcitedefaultseppunct}\relax
\EndOfBibitem
\bibitem[Torrie and Valleau(1980)Torrie, and Valleau]{Torrie:80}
Torrie,~G.~M.; Valleau,~J.~P. Electrical double layers. I. Monte Carlo study of
  a uniformly charged surface. \emph{J.~Chem. Phys.} \textbf{1980}, \emph{73},
  5807--5816\relax
\mciteBstWouldAddEndPuncttrue
\mciteSetBstMidEndSepPunct{\mcitedefaultmidpunct}
{\mcitedefaultendpunct}{\mcitedefaultseppunct}\relax
\EndOfBibitem
\bibitem[Leote~de Carvalho and Evans(1994)Leote~de Carvalho, and
  Evans]{LeotedeCarvalho1994}
Leote~de Carvalho,~R.; Evans,~R. The decay of correlations in ionic fluids.
  \emph{Mol. Phys.} \textbf{1994}, \emph{83}, 619–654\relax
\mciteBstWouldAddEndPuncttrue
\mciteSetBstMidEndSepPunct{\mcitedefaultmidpunct}
{\mcitedefaultendpunct}{\mcitedefaultseppunct}\relax
\EndOfBibitem
\end{mcitethebibliography}
\providecommand{\latin}[1]{#1}
\makeatletter
\providecommand{\doi}
  {\begingroup\let\do\@makeother\dospecials
  \catcode`\{=1 \catcode`\}=2 \doi@aux}
\providecommand{\doi@aux}[1]{\endgroup\texttt{#1}}
\makeatother
\providecommand*\mcitethebibliography{\thebibliography}
\csname @ifundefined\endcsname{endmcitethebibliography}
  {\let\endmcitethebibliography\endthebibliography}{}

\end{document}